\documentclass[preprint,nopreprintline,3p]{elsarticle}
\usepackage{amssymb}
\usepackage{color}
\usepackage{url}
\usepackage{booktabs}
\usepackage{multirow}
\usepackage{amssymb}
\usepackage{amsmath}
\usepackage{bm}
\usepackage{xspace}
\usepackage{rotating}
\usepackage{subfigure}

\journal{Biomedical Signal Processing and Control}

\begin{document}

\begin{frontmatter}

\title{A Deep Bayesian Neural Network for Cardiac Arrhythmia Classification with Rejection from ECG Recordings}

\author[1]{Wenrui Zhang}
\author[2]{Xinxin Di}
\author[3]{Guodong Wei}
\author[3]{Shijia Geng}
\author[3,4]{Zhaoji Fu}
\author[5,6]{Shenda Hong\corref{cor}}\ead{hongshenda@pku.edu.cn}

\address[1]{Department of Mathematics, Zhejiang University, Hangzhou, 310027, China.}
\address[2]{The First Affiliated Hospital of University of Science and Technology of China, Hefei, 230001, China}
\address[3]{HeartVoice Medical Technology, Hefei, 230027, China}
\address[4]{School of Management, University of Science and Technology of China, Hefei, 230026, China}
\address[5]{National Institute of Health Data Science at Peking University, Peking University, Beijing, 100191, China}
\address[6]{Institute of Medical Technology, Health Science Center of Peking University, Beijing, 100191, China}

\cortext[cor]{Corresponding author.}

\begin{abstract}
With the development of deep learning-based methods, automated classification of electrocardiograms (ECGs) has recently gained much attention. Although the effectiveness of deep neural networks has been encouraging, the lack of information given by the outputs restricts clinicians' reexamination. If the uncertainty estimation comes along with the classification results, cardiologists can pay more attention to ``uncertain'' cases. Our study aims to classify ECGs with rejection based on data uncertainty and model uncertainty. We perform experiments on a real-world 12-lead ECG dataset. First, we estimate uncertainties using the Monte Carlo dropout for each classification prediction, based on our Bayesian neural network. Then, we accept predictions with uncertainty under a given threshold and provide ``uncertain" cases for clinicians. Furthermore, we perform a simulation experiment using varying thresholds. Finally, with the help of a clinician, we conduct case studies to explain the results of large uncertainties and incorrect predictions with small uncertainties. The results show that correct predictions are more likely to have smaller uncertainties, and the performance on accepted predictions improves as the accepting ratio decreases (i.e. more rejections). Case studies also help explain why rejection can improve the performance. Our study helps neural networks produce more accurate results and provide information on uncertainties to better assist clinicians in the diagnosis process. It can also enable deep-learning-based ECG interpretation in clinical implementation.
 
\end{abstract}


\end{frontmatter}

\section{Introduction}
Recently, deep learning methods have achieved promising results in many electrocardiogram (ECG) applications \cite{hong2020opportunities, hannun2019cardiologist,ribeiro2020automatic,van2020automatic}, such as cardiac arrhythmia classification \cite{saman2019cardiac,clifford2017af,hong2020cardiolearn}, annotation \cite{peimankar2019ensemble}, sleep staging \cite{li2018method}, biometric identification \cite{labati2019deep,hong2020cardioid}, and cardiovascular health management \cite{fu2021artificial}. In contrast to traditional ECG analysis methods that compromise a two-stage strategy (``raw data'' to ``engineered features'' to ``predictions''), deep learning methods are in an end-to-end manner (``raw data'' to ``predictions'') by learning inherent representations from large-scale raw data directly. Computerized interpretation of ECGs is of increasing importance in clinical decision-making. However, reexamining the computer-assisted diagnosis of ECGs remains indispensable.

% limitation 
The research advances in deep learning methods on ECG data are developing more powerful deep neural network architectures for extracting more effective ECG representations, which are usually more complicated deep neural networks. Consequently, deep learning models are becoming increasingly complex. However, while much attention has been paid to the effectiveness of deep learning networks, a vital challenge is to be addressed before deep learning-based methods can be conducted in clinical practice. Owing to the increasing number of ECG recordings to be diagnosed, manual reexamination of automated ECG interpretation requires more information, such as the degree to which the model is sure about the outputs, rather than only the predictions from outputs of deep learning models. With more helpful information, cardiologists can arrange their time in different cases more accurately. It is highly expected that predictions from neural networks are the most certain ones, so the remaining cases can be paid more attention by experienced cardiologists. 

% a potential solution
Classification with rejection is a potential solution to this problem. Classifiers can choose not to make classification predictions when it is not ``certain'' about the prediction  to avoid critical misclassifications and leave the equivocal cases to be diagnosed by clinicians. Although the Softmax outputs of a neural network can be used as the criterion to reject predictions, 
they sometimes result in incorrect predictions with a high probability of unseen data \cite{louizos2017multi}. Due to the seriousness of medical diagnosis, the outputs of Softmax are not suitable in clinical scenarios. Therefore, we propose a classification using a rejection solution. Our rejection is based on two types of uncertainties \cite{kiureghian2009aleatory}, the data uncertainty and the model uncertainty, which are due to the noise in data and the lack of data, respectively. These two uncertainties are associated with the outputs of Softmax and can be regarded as more robust measures for the degree of confidence in predictions.

% contributions
In our study, we aim to achieve classification with rejection from ECG recordings, based on data uncertainty and model uncertainty, using deep learning techniques. In detail, we first constructed a very deep Bayesian neural network with 61-layer convolutional layers to classify cardiac arrhythmias from ECG recordings. Then, we conducted multiple tests using our trained model with dropout. The model uncertainty and data uncertainty were computed using the predicted probabilities from multiple tests. Finally, the classification with rejection is made by determining whether the sum of the two uncertainties meets a predefined uncertainty threshold, which is set according to the actual conditions. The results prove that our method can help discard a larger proportion of incorrect predictions, yielding higher metrics in accepted predictions compared with accepting all predictions.

\section{Materials and Methods}

\subsection{Dataset}

We used the training set of real-world 12-lead ECG recordings from the 2018 China Physiological Signal Challenge (CPSC) \cite{liu2018open} \footnote{\url{http://2018.icbeb.org/Challenge.html}}. It contains 6,877 (3,699 male, 3,178 female) 12-lead ECG recordings lasting in duration from 6 s to 60 s, collected from 11 hospitals sampled at 500 Hz. Of these, 918 recordings were normal (normal), 1,098 recordings were atrial fibrillation (AF), 704 recordings were first-degree atrioventricular block (AVBI), 207 recordings were left bundle branch block (LBBB), 1,695 recordings were right bundle branch block (RBBB), 556 recordings were premature atrial contraction (PAC), 672 recordings were premature ventricular contraction (PVC), 825 recordings were ST-segment depression (STD), and 202 recordings were ST-segment elevated (STE). Our task is to classify cardiac arrhythmia cases among them.

\subsection{A Deep Bayesian Neural Network for Modeling ECG Data}

A 61-layer deep Bayesian neural network convolutional neural network (CNN) is designed to model ECG recordings. The detailed model architecture is presented in Table \ref{tb:model}. Overall, we follow the recent state-of-the-art model architecture for image classification, which deploys a neural architecture space search strategy to find a family of best models \cite{radosavovic2020designing}, but replace the filter shape from 2-dimensional patches to 1-dimensional strips. Our modified network consists of seven stages, which contain 2,2,2,3,3,4,4 blocks in each stage. Blocks are residual-connected with shortcut connections \cite{he2016deep,he2016identity}. Each block is a bottleneck architecture - a cascade of one convolutional layer (Conv1) with kernel size set to 1, one aggregated convolutional layer (ConvK) \cite{xie2017aggregated} with kernel size set to 16, groups set to 16, and one convolutional layer (Conv1) with kernel size set to 1. In each stage, the first block down-sampled the input length (last dimension) by a factor of 2. Meanwhile, the corresponding shortcut connections down-sample the identity input using a max pooling operation by a factor of 2. Before each convolution layer, there is a nonlinear transformation, which is a combination of batch normalization (BN) \cite{ioffe2015batch}, Swish activation \cite{ramachandran2017searching}, and dropout \cite{srivastava2014dropout}. To further improve the model performance, we also introduce a channel-wise attention mechanism (SE block) \cite{hu2018squeeze}. After seven stages of convolutional layers, the output is reduced by a global average pooling layer. The prediction layer is a fully connected, dense layer that generates logits.

\begin{table}[h]
\centering
\begin{tabular}{lll}
\toprule
Stage      & Layers                    & Output size   \\
\midrule
Input      &                           & (*, 12, 5000) \\
Stage 1    & (Conv1, ConvK, Conv1) x 2 & (*, 64, 2500) \\
Stage 2    & (Conv1, ConvK, Conv1) x 2 & (*, 160, 1250) \\
Stage 3    & (Conv1, ConvK, Conv1) x 2 & (*, 160, 625) \\
Stage 4    & (Conv1, ConvK, Conv1) x 3 & (*, 400, 312) \\
Stage 5    & (Conv1, ConvK, Conv1) x 3 & (*, 400, 156)  \\
Stage 6    & (Conv1, ConvK, Conv1) x 4 & (*, 1024, 78) \\
Stage 7    & (Conv1, ConvK, Conv1) x 4 & (*, 1024, 39) \\
Pooling    & Global Average            & (*, 1024)     \\
Prediction & Dense                     & (*, 9)        \\
\bottomrule
\end{tabular}
\caption{Model architecture. $n=2000$, $d=12$, $c=3$. The first dimension $*$ represents the number of samples in a batch. }
\label{tb:model}
\end{table}

Formally, we use $x \in \mathbb{R}^{d \times n}$ to represent the input ECG data, where $n$ is the length of the ECG and $d$ is the number of leads, which is 12 in our case. We also use $\mathcal{F}$ to represent the deep neural network. Then, the predicted logits $z \in \mathbb{R}^{c}$ can be represented as
\begin{equation}
z = \mathcal{F}(x)
\end{equation}
After applying \textit{Softmax} to $z$, we obtain the probability $y \in \mathbb{R}^{c}$.

Dropout was initially used to prevent overfitting; however, in our study, we used it to sample an approximated posterior distribution. Before each layer, we applied dropout to discard the units in this layer with probability $p$. Thus, although feeding with identical inputs, the model with dropout can produce different outputs. Due to the dropout before each layer, our neural network is mathematically equivalent to an approximation of the probabilistic deep Gaussian process \cite{damianou2013deep}.  
\begin{figure}
\centering
\includegraphics[width=0.5\linewidth]{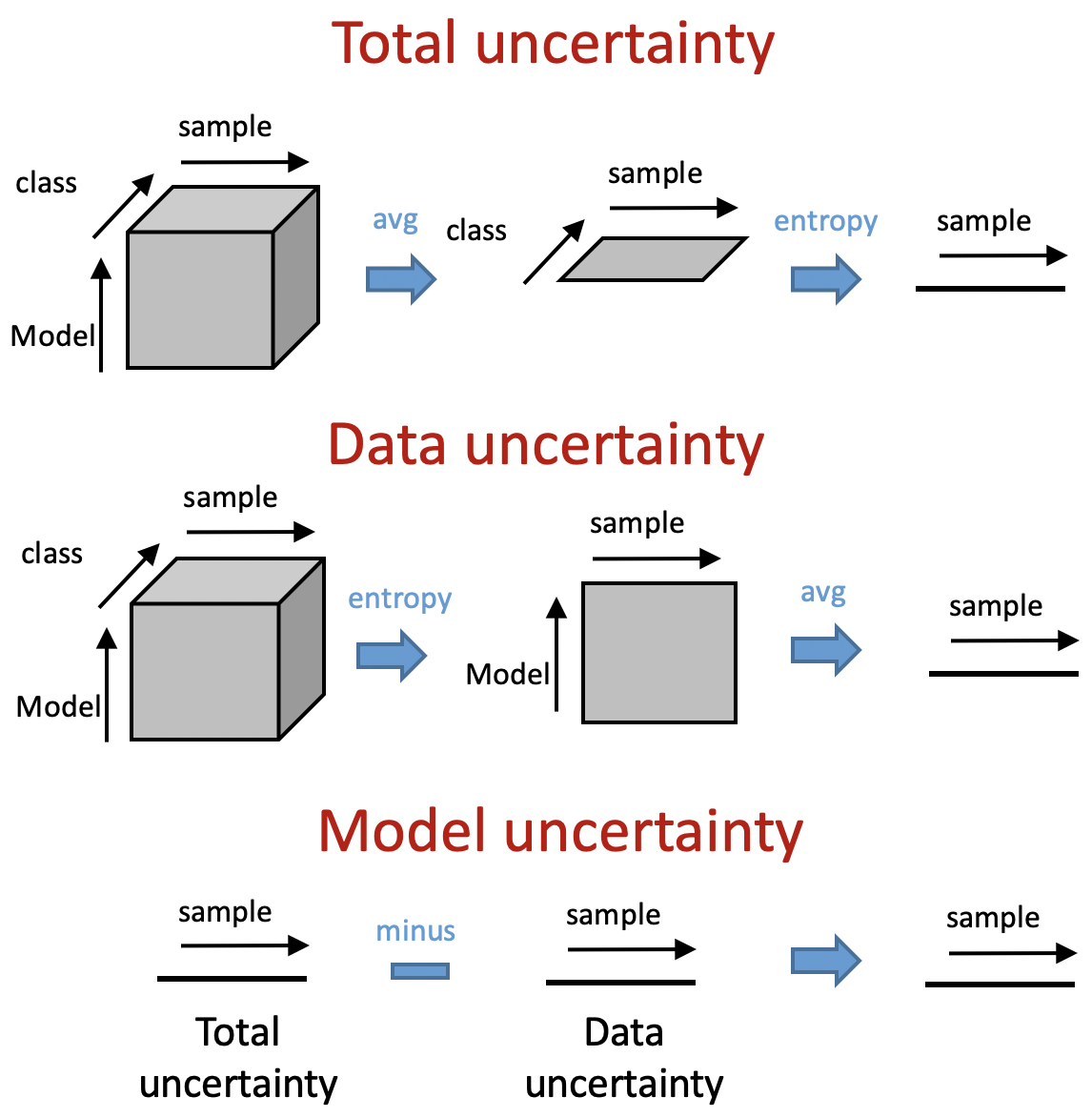}
\caption{ Diagrams of computing model uncertainty and data uncertainty.  }
\label{fig:compute}
\end{figure}

\subsection{Model Uncertainty and Data Uncertainty}
There are two main types of uncertainties in the Bayesian neural network: data uncertainty (\textit{aleatory} uncertainty) and model uncertainty (\textit{epistemic} uncertainty). Data uncertainty could be related to the noise inherent in observations, which is irreducible with more data, for example, imprecision in measurement. Data uncertainty can be further classified as \textit{homoscedastic} uncertainty, which is consistent among various inputs, and \textit{heteroscedastic} uncertainty, which depends on the inputs. Model uncertainty represents the uncertainty in model parameters or \textit{structure} uncertainty (which model we choose) and can be reduced by increasing the size of the training dataset. Measuring the two types of uncertainties enables us take to actions to improve model performance \cite{malinin2018predictive}.

To obtain a comprehensive estimate of uncertainty, we calculated the sum of the two uncertainties (referred to as the total uncertainty). We train our classification model $\mathcal{F}$ on the training set $\mathcal{D} = \{x_i, y_i\}$ ($y_i\in\{0,1,\cdots, K\}$). During testing, we keep dropout on, and conduct tests for N = 50 times to generate slightly different predictions {$\hat{y_i}, i = 1, 2, \cdots, N$}. To quantify the total uncertainty, we computed the predicted average across all models and calculated the sample-wise predictive entropy:
\begin{equation}
\mathrm{Total\ Uncertainty} = \mathcal{H}[\mathbb{E}_i[\hat{y_i}]] = \mathcal{H}[\frac{\sum_{i=1}^N\hat{y_i}}{N}]
\end{equation}
where $\mathcal{H}[\mathrm{\hat{y}}]$ is the entropy of predictive probabilities:
$$
\mathcal{H}[\hat{y}] = - \sum^K_{c=1}\hat{y^c}\ln(\hat{y^c})
$$
where $\hat{y^c}$ represents the predictive probability and the corresponding $x$ belongs to class $c$. To compute data uncertainty, we first compute the sample-wise predictive entropy and then take the average across all predictions.
\begin{equation}
\mathrm{Data\ Uncertainty} = \mathbb{E}_i [\mathcal{H}[\hat{y_i}]] = -\frac{\sum_{i=1}^N(\sum_{c=1}^K\hat{y_i^c}\ln (\hat{y_i^c}))}{N}
\end{equation}
Then, the model uncertainty can be calculated by subtracting the data uncertainty from the total uncertainty. 
\begin{equation}
\mathrm{Model\ Uncertainty} = \mathrm{Total\ Uncertainty}  - \mathrm{Data\ Uncertainty} 
\end{equation}
The diagram of computing uncertainties is shown in Figure \ref{fig:compute}.

\subsection{Prediction with Rejection based on Uncertainties}
For each data point in the test set, the data uncertainty was calculated. When the uncertainty of one data point is too high, we can assume that the trained model is ``uncertain'' about this data point. Consequently, we can set a threshold $t$, which is relative to the uncertainties of the entire test set or determined by other conditions. If the uncertainty is greater than $t$, we choose not to make a prediction. On the contrary, if the uncertainty is small enough, we have more confidence in our prediction and choose to adopt it.

\subsection{Implementation Details}

In model training, we follow the original data split, which separates the entire dataset into 80\% training set, 10\% validation set, and 10\% test set, by subjects. Hyperparameters were selected on the validation set. The results are reported for the test set. In addition, we added a weight norm to avoid overfitting. The Adam \cite{kingma2014adam} optimizer with back-propagation was used to train the model. The batch size was set to 256. The learning rate was set to 0.001 initially and then reduced by a factor of 0.3 when the validation performance stopped improving in 6000 steps. The ordinal loss is in the form of multi-task learning, which makes it more difficult to train than conventional classification tasks. To improve this, we first trained the neural network with conventional cross-entropy loss, then replaced the objective with ordinal loss and fine-tuned the weights. PyTorch was used to build and train the model. Our code is publicly available at \url{https://github.com/hsd1503/ecg_uncertainty}. 

In the test phase, we used the Macro-F1 score of the nine classes to evaluate the model performance. For each class, we set it as positive and the other classes as negative and calculated the F1-score of this class. Then, we calculated the average value of F1-scores for the nine classes.
\begin{figure}
\begin{minipage}[htb]{0.45\linewidth}
\includegraphics[width=\linewidth]{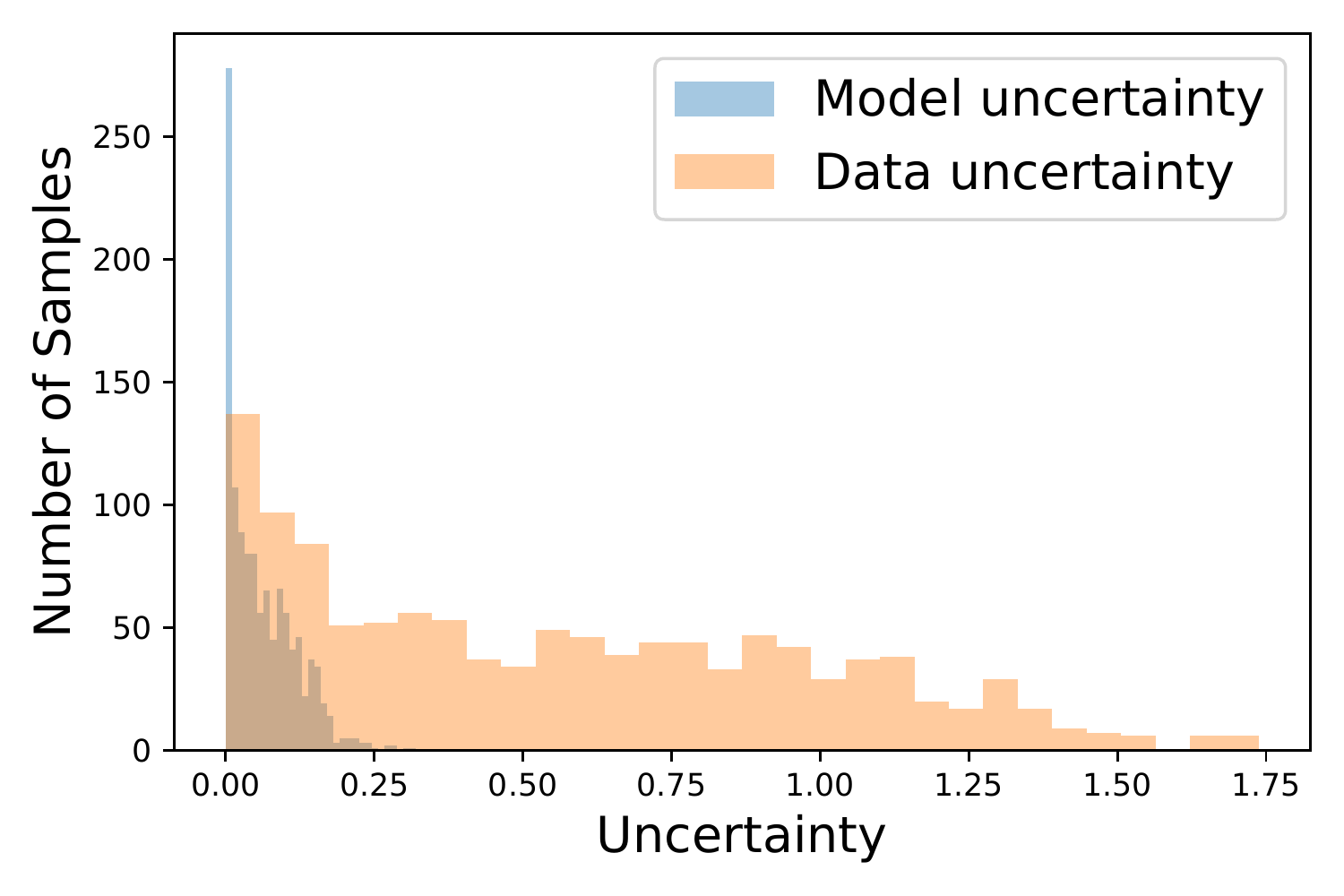}
\caption{Distributions of model uncertainty and data uncertainty.}
\label{fig:uncertainty_distribution}
\end{minipage}
\quad
\begin{minipage}[htb]{0.45\linewidth}
\includegraphics[width=0.9\linewidth]{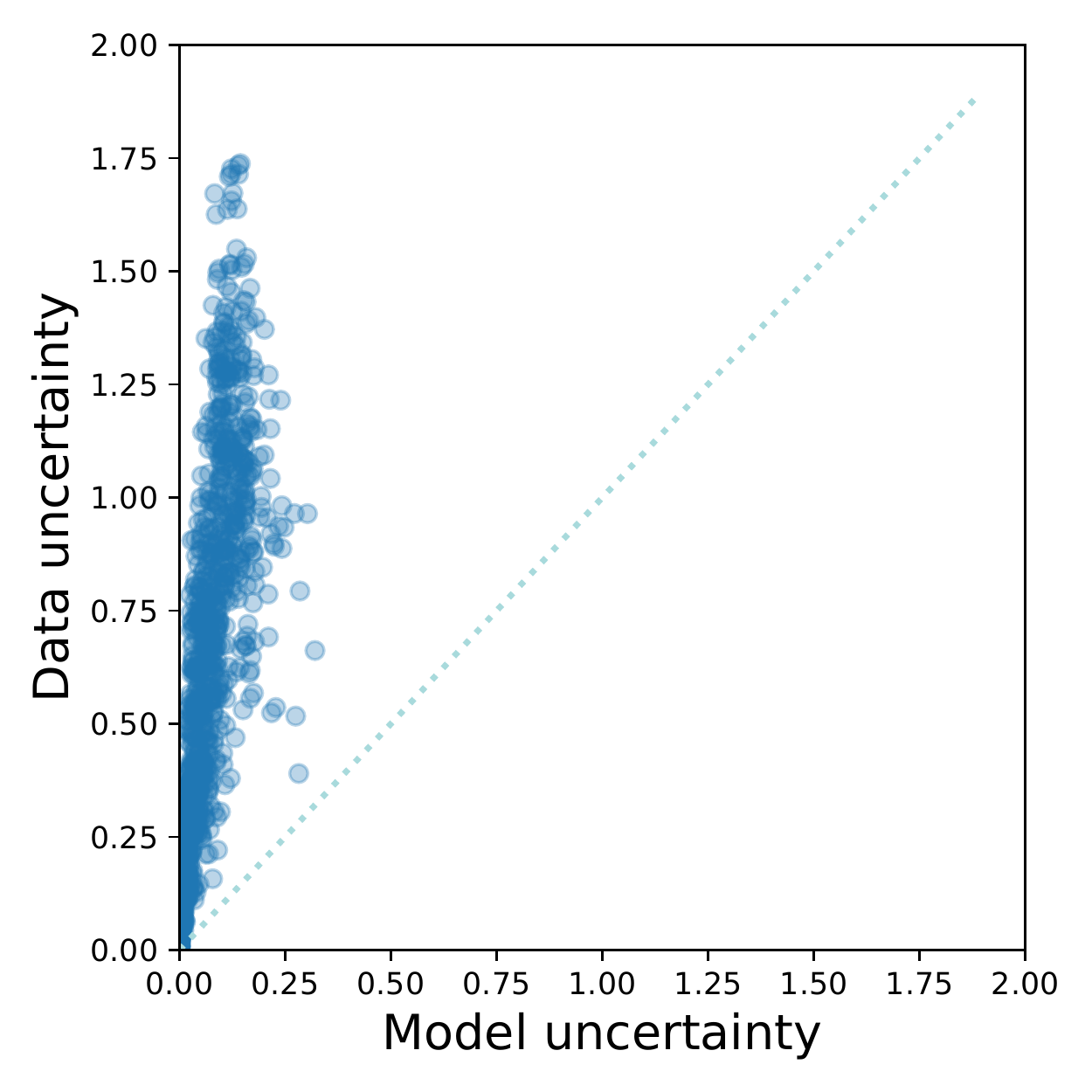}
\caption{Model uncertainty vs. data uncertainty.}
\label{fig:corr}
\end{minipage}
\end{figure}
\begin{figure}
    \centering
    \subfigure[AVBI]{
    \includegraphics[scale = 0.5]{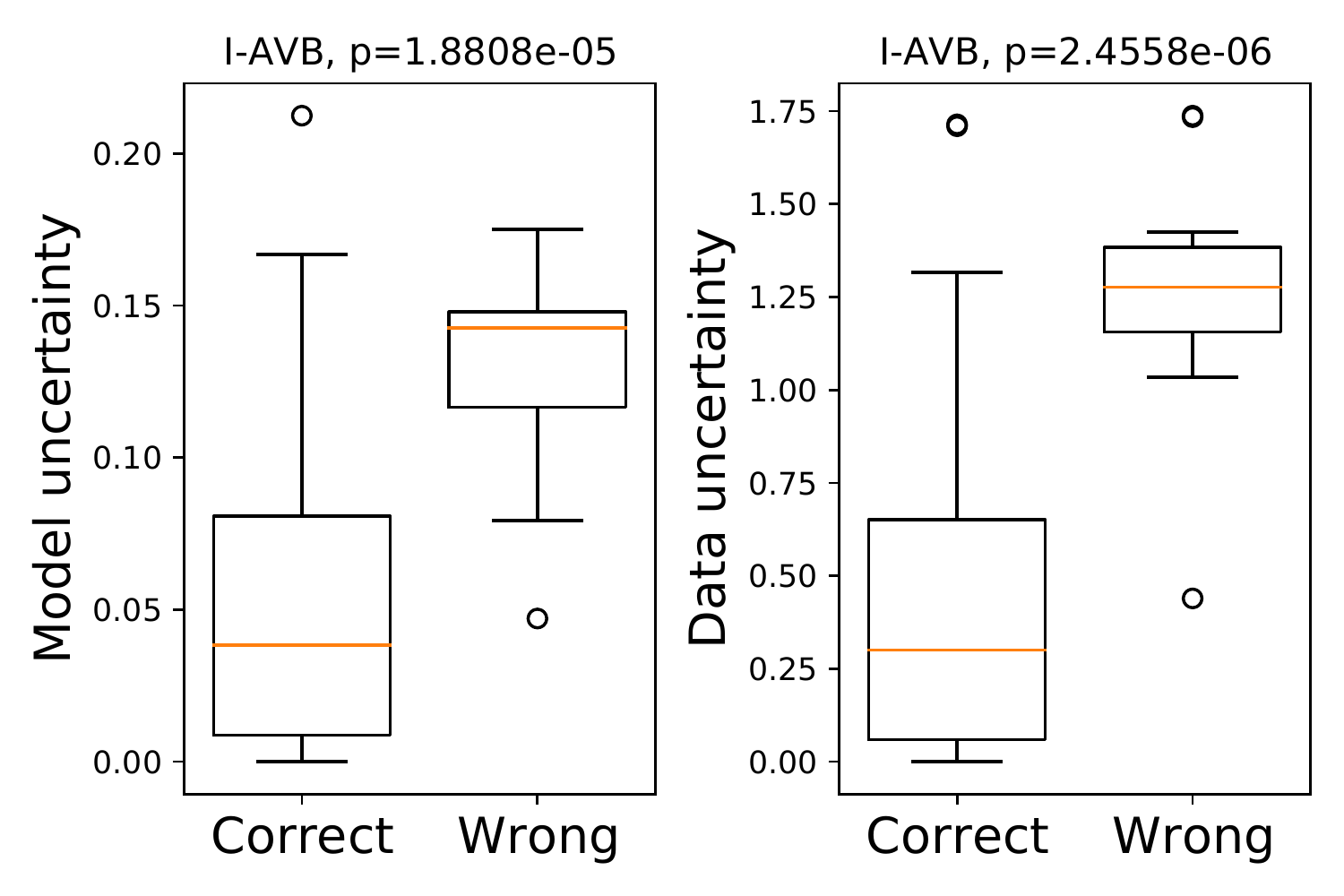}
    }\subfigure[STE]{
    \includegraphics[scale = 0.5]{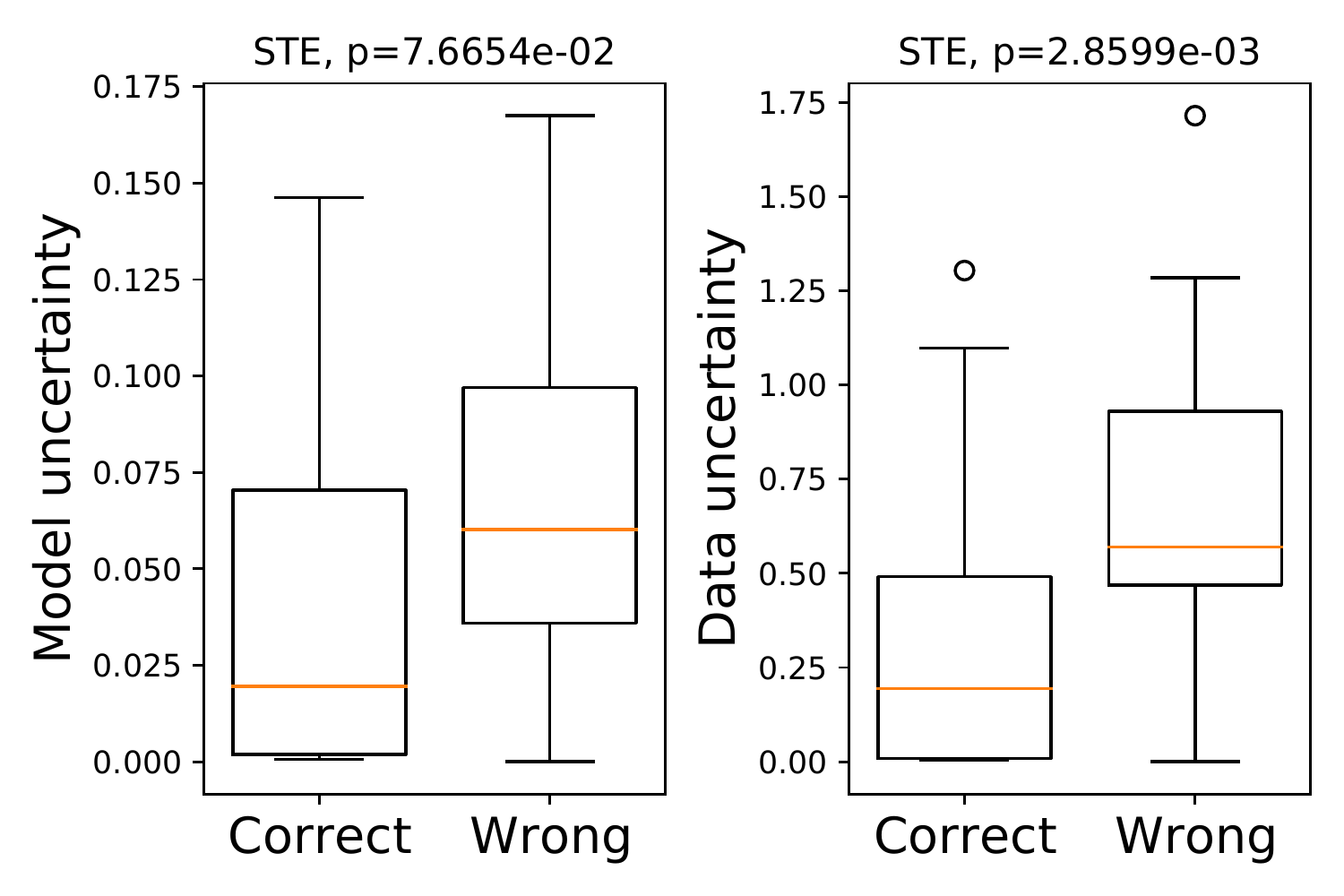}
    }
    \caption{The distributions of data uncertainty and model uncertainty of correctly and wrongly classified samples of AVBI and STE.}
    \label{fig:comparison}
\end{figure}

\section{Results}
\subsection{Relationship of Uncertainties}

% which one is bigger
We calculated two types of uncertainties, model and data, for each sample and grouped the samples based on the uncertainties. Figure \ref{fig:uncertainty_distribution} shows the distributions of the model uncertainty and data uncertainty. The model uncertainty in our experiment was generally smaller, ranging from 0 to 0.25. However, the data uncertainty has a wider distribution and is likely to be higher than the model uncertainty. In addition, as shown in Figure \ref{fig:corr}, the uncertainty points are all above the line $y=x$, which indicates that all of the x-coordinates of points (model uncertainty) are smaller than the y-coordinates of points (data uncertainty). Therefore, the model uncertainty was also less than the data uncertainty for each sample.

% linear relationship
In addition, as shown in Figure \ref{fig:corr}, the data uncertainty and model uncertainty, in our experiment, are positively correlated. Moreover, we analyzed the Pearson correlation for both uncertainties in each class. Overall, the Pearson correlation coefficient was 0.7857 and the p-value for testing non-correlation was 4.8026e-245; thus, we can conclude that the two types of uncertainties are strongly correlated. In addition, the uncertainty points are all above line $y=x$. Therefore, the model uncertainty was less than the data uncertainty for all samples in our test set.

For each class of disease, the data uncertainty and model uncertainty are also distributed unevenly. In Figure \ref{fig:comparison}, we present two examples of correctly and incorrectly classified samples, with both uncertainties for AVBI and STE (the diseases with the best and worst prediction performances). We also performed Welch's t-test to determine uncertainties for both correctly and incorrectly classified samples. In terms of each type of uncertainty and each class of disease, the null hypothesis is that the average values of uncertainties for correctly classified samples and wrongly classified samples are identical. The alternative hypothesis is that the uncertainty of correctly classified samples is greater than that of incorrectly classified ones. Except for the model uncertainty of PAC, all p-values are smaller than 0.1, and most of them are smaller than 0.01. According to the p-values of t-tests for each class, and all classes, we conclude that samples with less uncertainty are more likely to be classified correctly.

\begin{table}[]
\centering
\begin{tabular}{lccccccc}
\toprule
\multirow{2}{*}{Label} & Model        & Data       & \multirow{2}{*}{Performance}       & Data & Model & Correlation & Correlation\\
                       &Uncertainty  & Uncertainty &  & p-value& p-value& coefficient& p-value           \\
\midrule
Normal                 & 0.6119 & 0.0624            & 0.6538      & 3.10e-06***       & 2.45e-05***  &0.7278 & 6.66e-19***     \\
AF                     & 0.3599& 0.0332            & 0.7593      & 1.08e-05***       & 5.68e-05***   &0.8900 & 4.27e-05***     \\
AVBI                    & 0.5205 & 0.0624          & 0.8878      & 2.46e-06***       & 1.88e-05***  &0.8864 & 7.14e-34***      \\
LBBB                   & 0.4477& 0.0269            & 0.7037      & 7.55e-05***       & 1.58e-06***  &0.8709&2.14e-11***  \\
RBBB                  & 0.3116 & 0.0277            & 0.8577      & 7.67e-07***       & 8.98e-07***  &0.8369 & 4.18e-67***      \\
PAC                    & 0.8198& 0.1001            & 0.4931      & 3.53e-02** \ \     & 0.395       &0.5319 & 9.97e-14*** \\
PVC                    & 0.6355& 0.0828            & 0.5731      & 2.42e-08***       & 1.63e-08***  &0.7137 & 1.43e-30***      \\
STD                    & 0.7199 & 0.0763           & 0.6326      & 7.84e-28***       & 6.62e-14***  &0.8258 & 3.81e-38***      \\
STE                    & 0.5617  & 0.0592          & 0.4103      & 2.86e-03***      & 7.67e-02*\ \ \ \ & 0.7553 & 5.54e-10***         \\
\midrule
Overall & 0.5538 & 0.0612 &0.6635 &6.55e-62*** & 1.94e-48*** & 0.7857 & 4.80e-245***\\
\bottomrule
\end{tabular}
\caption{ Compare uncertainties, performances, and p-values. The number of * represents the significance level, where * means significance level at 0.1, ** means significance level at 0.05, and *** means significance level at 0.01. }
\label{tb:perf}
\end{table}

\subsection{ Results of Prediction with Rejection  }
 
\begin{figure}
\centering
\includegraphics[width=0.7\linewidth]{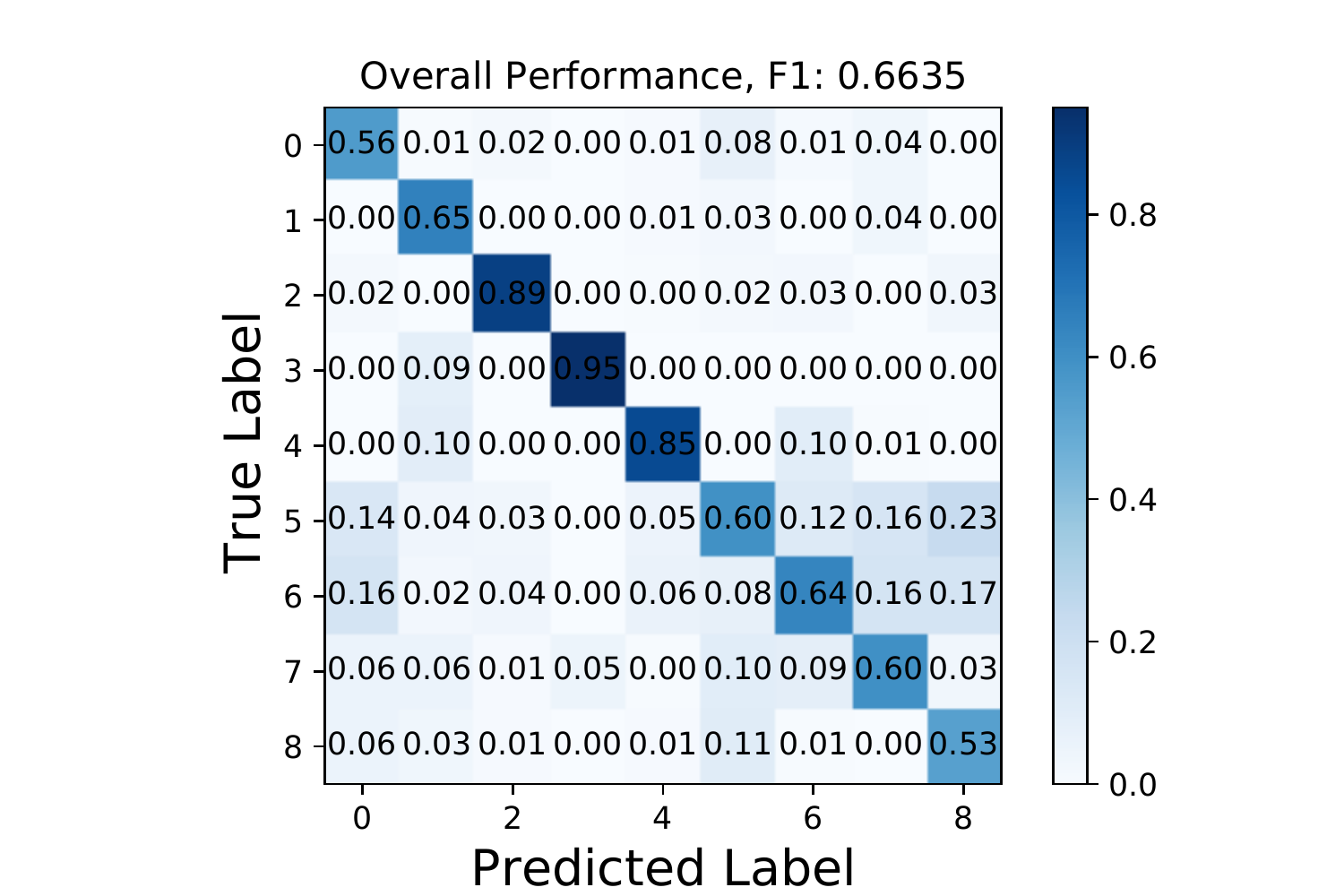}
\caption{The confusion matrix of predictions without rejection. }
\label{fig:overall_cm}
\end{figure}

\begin{figure}
\centering
\includegraphics[width=0.7\linewidth]{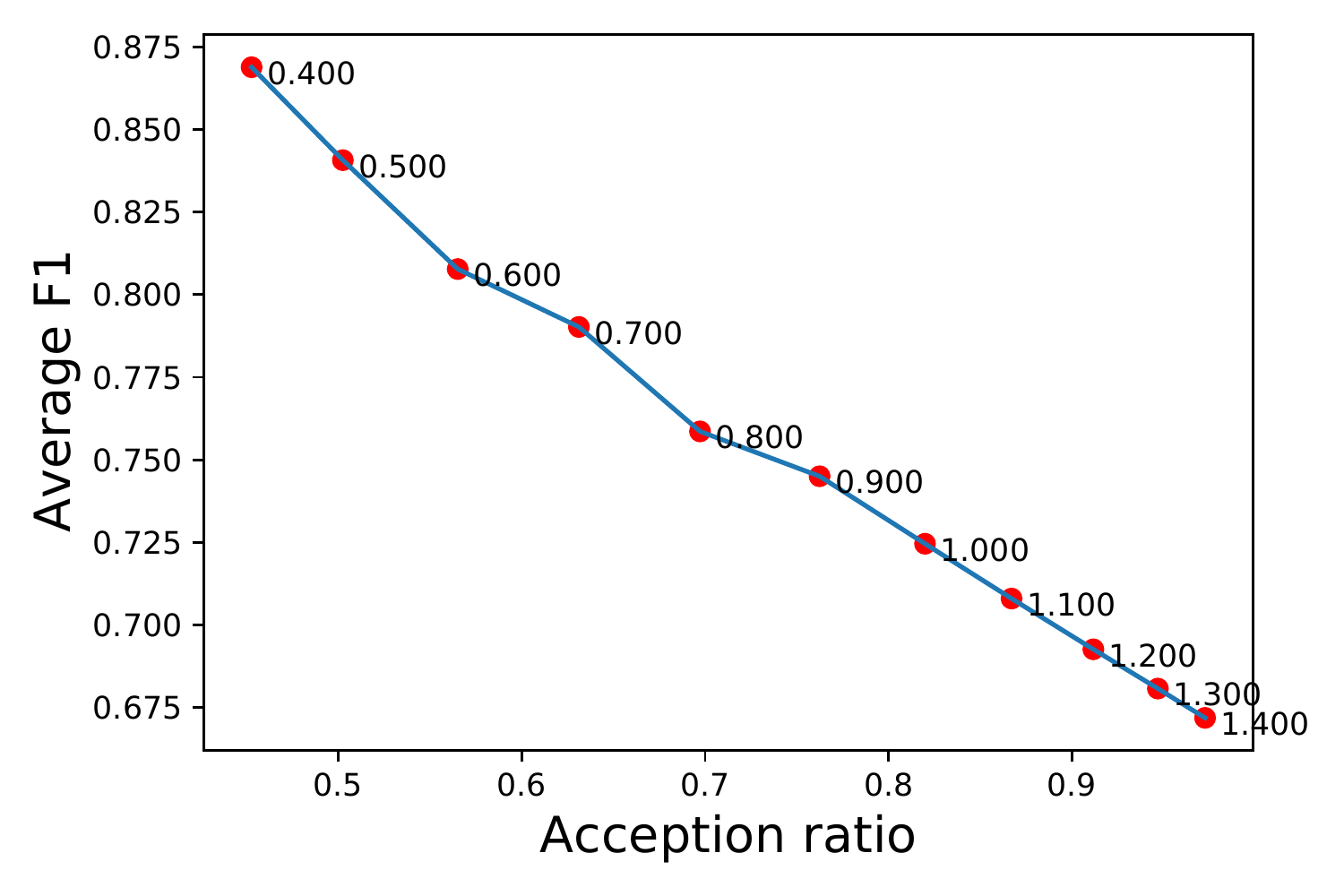}
\caption{  Model performance on different rejection thresholds.  }
\label{fig:ratio}
\end{figure}

We show the precision confusion matrix (normalizing each row using the sum of each row) of predictions without rejection in Figure \ref{fig:overall_cm}. The average F1-score of the nine classes was 0.6635. In addition, we test the predictions of our model with rejection under different thresholds, as shown in Figure \ref{fig:ratio}. The values near the points are the thresholds, and the x -and y-axes are the accept ratios corresponding to thresholds and average F1-score, respectively. The thresholds range from 0.400 to 1.500 with an interval of 0.050, and the acceptance ratios ranged from 0.453 to 0.979. F1-scores are negatively correlated with acceptance ratios; thus, performance is positively correlated to thresholds. The largest F1-score is 0.8688 with a threshold of 0.400, which is 0.2053 larger than the F1-score without rejection.

\begin{figure}
\centering
\includegraphics[width=0.85\linewidth]{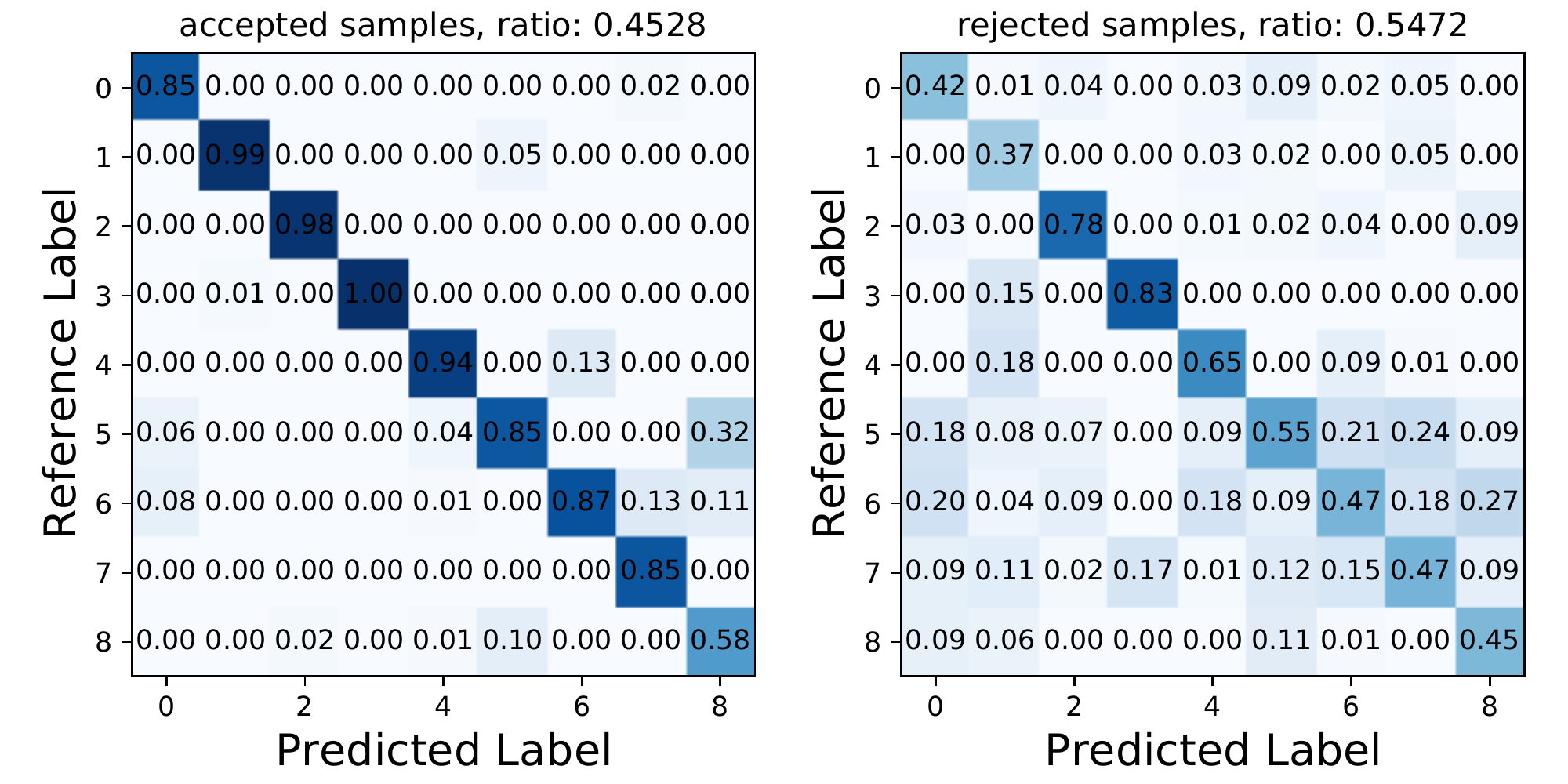}
\caption{  Confusion matrix on accepted samples (left) and rejected samples (right).  }
\label{fig:groups}
\end{figure}

The precision confusion matrices for the accepted and rejected samples, under the threshold of 0.400, are shown in Figure \ref{fig:groups}. In terms of the accepted samples, 0.4528 predictions were accepted, and all precision values of the nine classes improved compared with Figure \ref{fig:overall_cm}. Predictions in the four classes can yield a precision of more than 0.90, and most precision values are not less than 0.85. Correspondingly, the remaining 0.5472 predictions yielded smaller precision values. The precision values of all classes were less than 0.85, and most of them were smaller than 0.60. By comparing the two figures, we believe that our model is more certain for samples with small uncertainty, and rejecting can raise the proportion of correctly classified samples.

\subsection{ Case Studies}
We also conducted some case studies. From the samples that were incorrectly predicted, 30 samples with the largest data uncertainties and 30 samples with the smallest data uncertainties were selected, and we looked for the reasons why the former ones have large data uncertainties, and why the latter ones are incorrectly predicted. One experienced cardiologist reexamined the results and attempted to explain the possible reasons medically. 

\begin{figure}
    \centering
    \subfigure{
    \includegraphics[scale = 0.2]{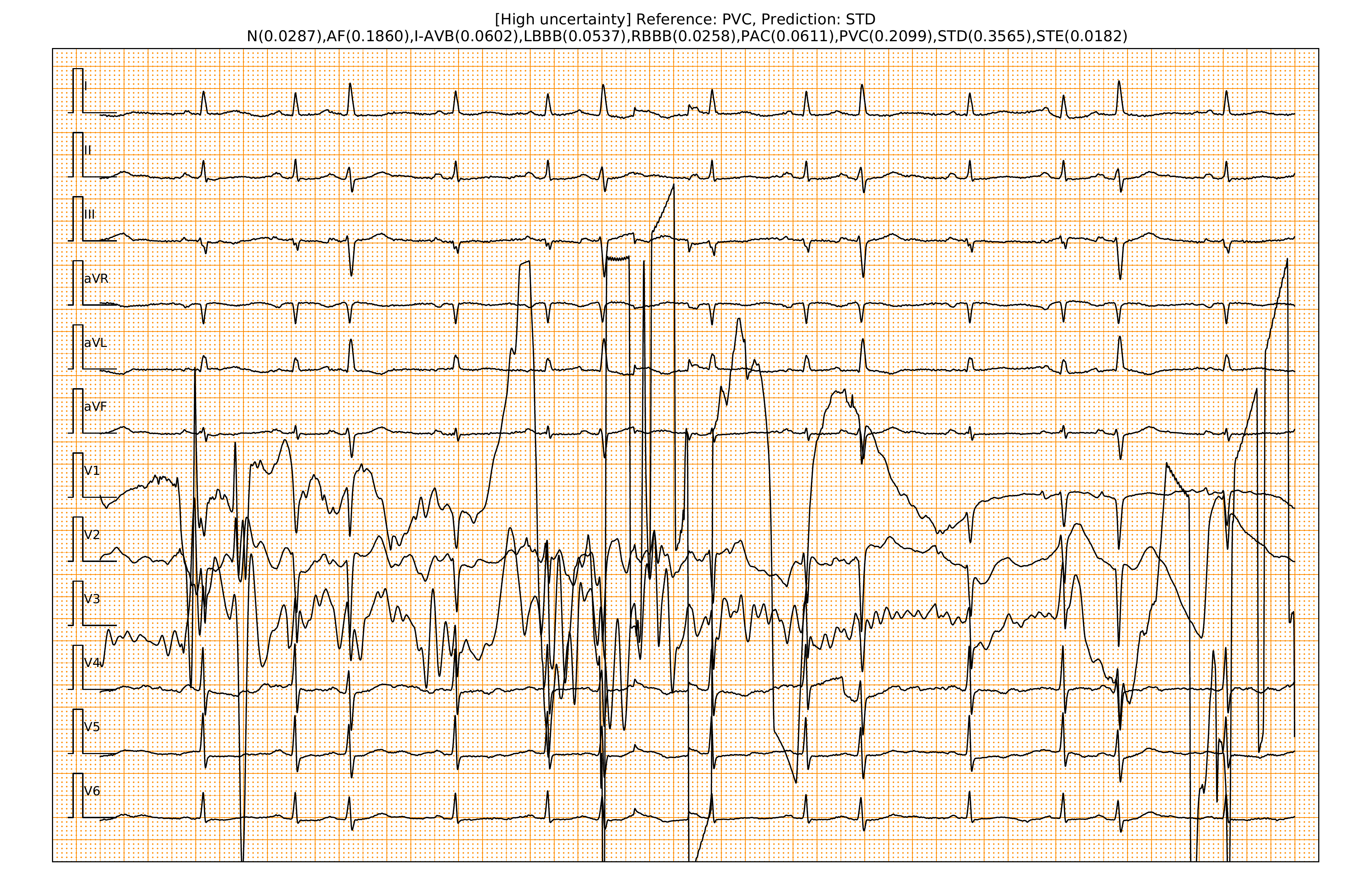}
    }
    \subfigure{
    \includegraphics[scale = 0.2]{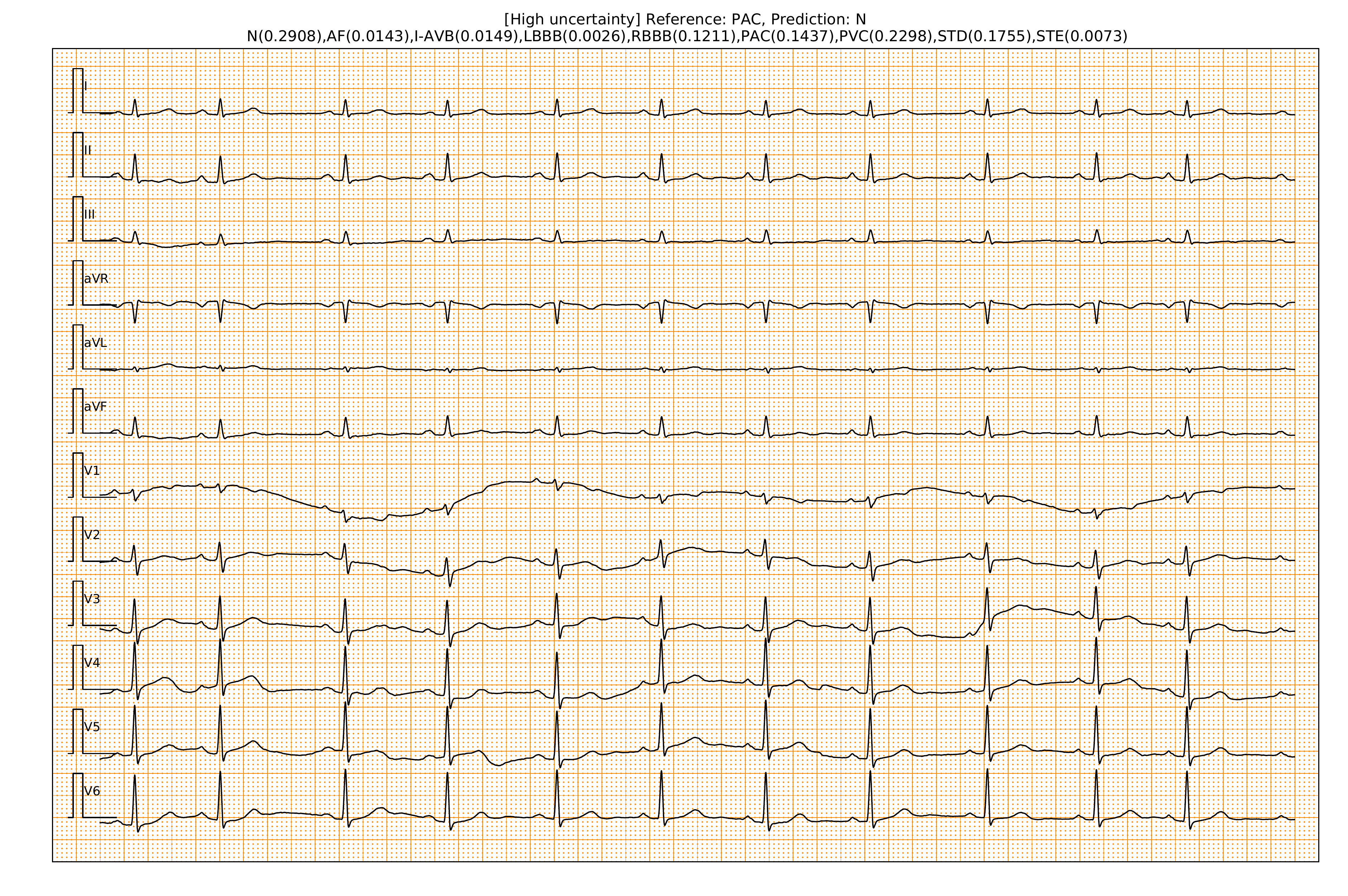}
    }
    \caption{Two cases with high data uncertainties}
    \label{fig:high_noise}
\end{figure}

In terms of samples with the largest data uncertainties, the cardiologist explained that the large data uncertainties were mainly caused by data noise. Data noise can be categorized as follows:
\begin{itemize}
\item \textbf{Large interference}. Some recordings are accompanied by large interference, which makes classification difficult. For example, as shown in Figur \ref{fig:high_noise}, there are some large waves in some leads. 
\item \textbf{Baseline drift}. Baseline drift is also one of the most common types of noise in ECG data. As shown in Figure \ref{fig:high_noise}, the baseline drift occurs in some leads, making the average value of data change irregularly. Thus, these data points are outside the distribution of the normal training data.
\end{itemize}

\begin{figure}
    \centering
    \subfigure{
    \includegraphics[scale = 0.2]{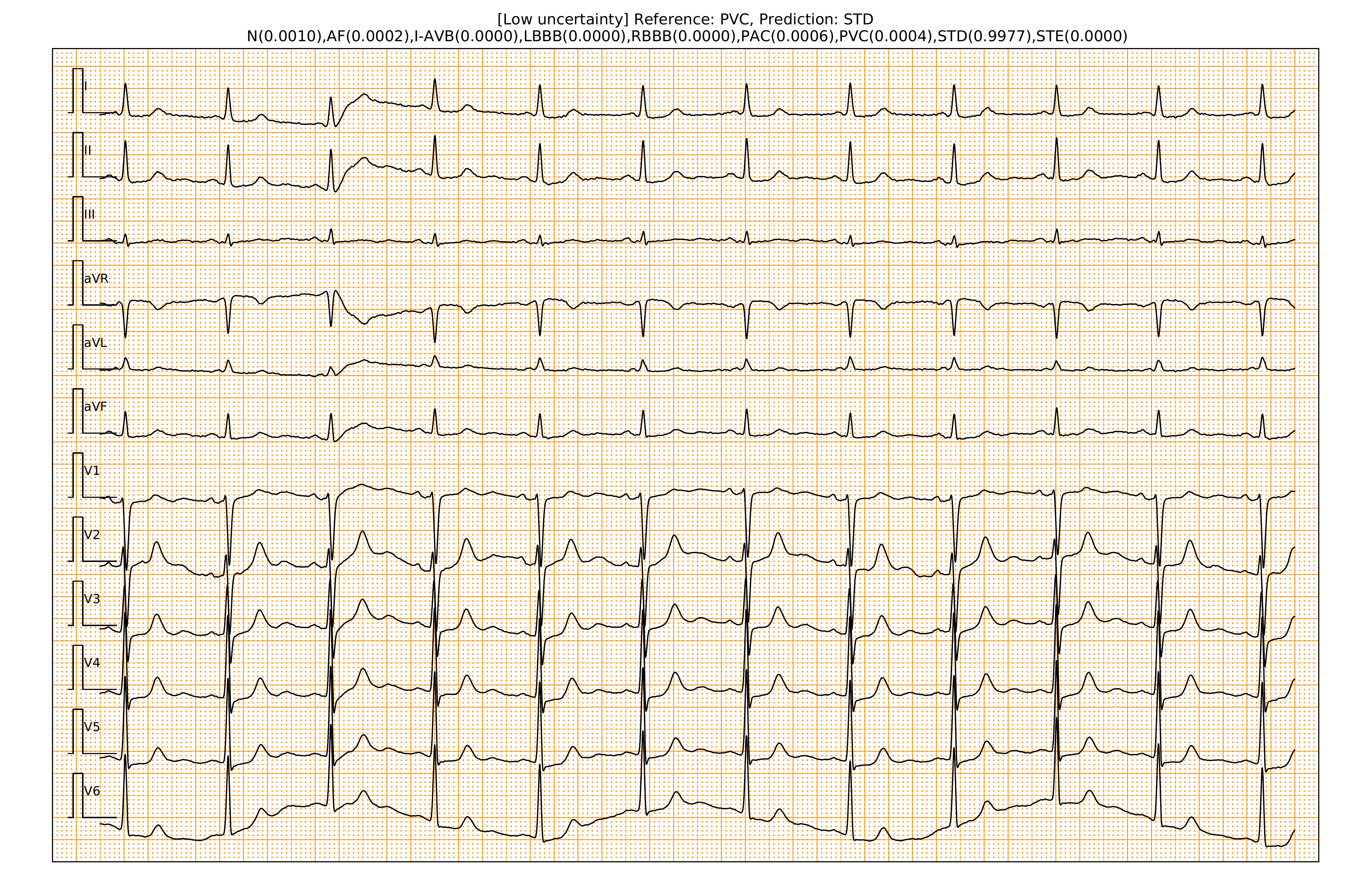}
    }
    \subfigure{
    \includegraphics[scale = 0.2]{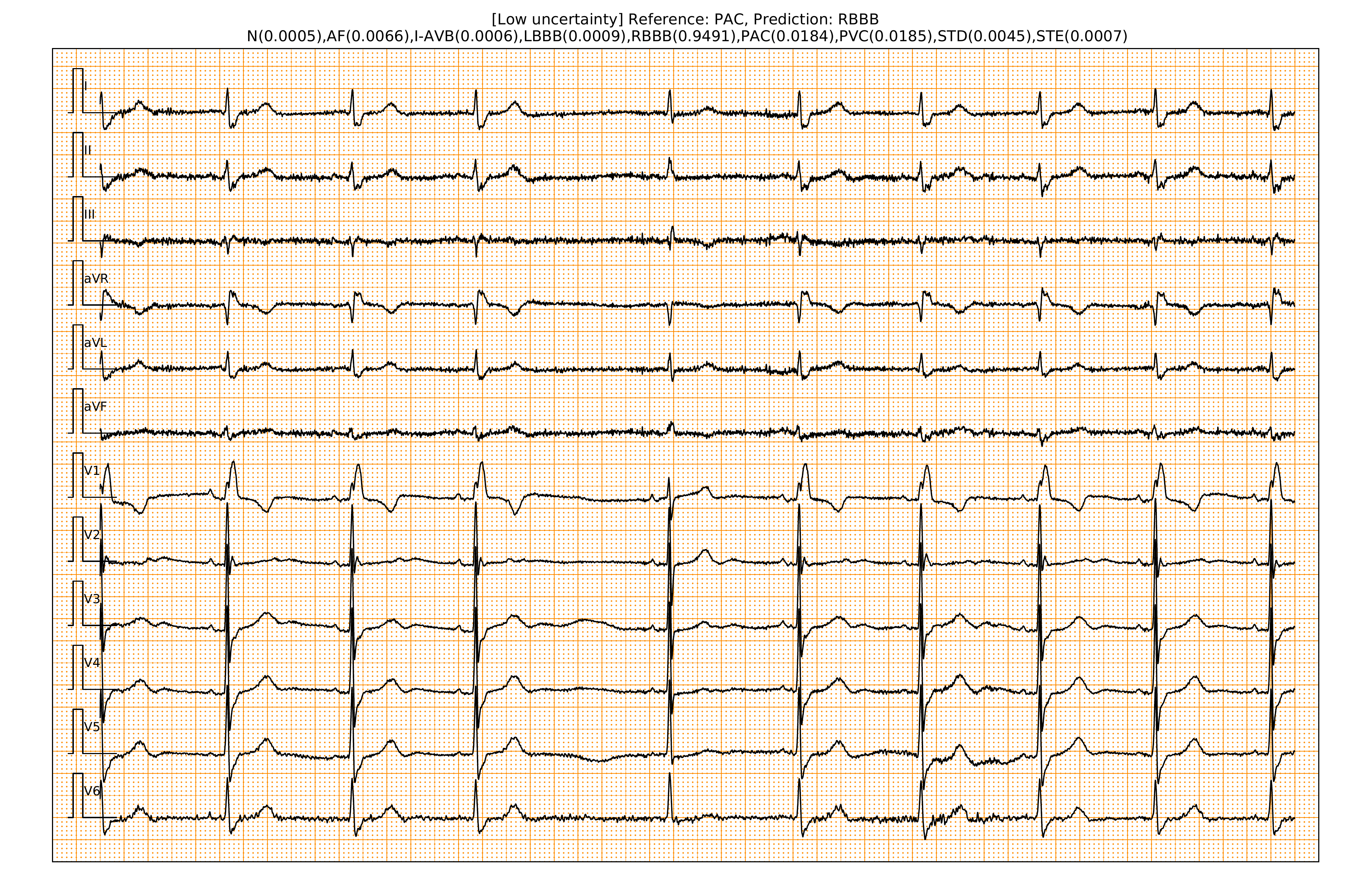}
    }
    \caption{Two cases with low data uncertainties but wrongly predicted}
    \label{fig:low_label}
\end{figure}

For samples with the smallest data uncertainties, the disagreement of labels accounts for most of the incorrect predictions. There are two types of label disagreement.
\begin{itemize}
    \item \textbf{Disagreement between experts}. For example, the left figure, shown in Figure \ref{fig:low_label}, is originally labeled as PVC, and the prediction of our model gives STD, while our cardiologist agrees more with STD classification.
    \item \textbf{Mixed labels}. Some samples appeared to be a mixture of the two diseases. Our model produces one of these, while the label is another. The label exists, but it is not complete. For example, the right figure, shown in Figure \ref{fig:low_label}, is thought to be both RBBB and PAC, but PAC was not conducted. Our predicted label was RBBB, but the original label was relatively insufficient.
\end{itemize}

According to the explanations given by cardiologists, we can find that samples with high data uncertainties are indeed difficult to classify, even for cardiologists, and that incorrectly predicted (correctly predicted in fact) samples with small uncertainty are more likely due to the disagreement of labels.

\section{Discussion}

The results show that prediction with rejection can improve the prediction quality. As shown in Figure \ref{fig:comparison} and Table \ref{tb:perf}, it is clear that the data uncertainties of correctly classified samples are less than those of wrongly classified samples, because the p-values are less than 0.01. From Figure \ref{fig:ratio}, we can conclude that rejected samples are more likely to be incorrectly predicted. Thus, as we reject more predictions, the proportion of correct predictions increases. By comparing Figure \ref{fig:overall_cm} with Figure \ref{fig:groups}, we can see that the precision values of all diseases improve, which means that our rejection can benefit the prediction of any disease.

To determine why rejection can improve the performance, we undertook case studies. Based on the case studies, we can conclude that large noise accounts for high data uncertainty. At the same time, large data noise means that it is difficult to classify the sample, even for cardiologists. Consequently, data uncertainty is regarded as positively correlated with classification difficulty. When we discard samples with high uncertainties, it is more possible for us to discard the ``difficult'' samples.

Our prediction with rejection can facilitate automated detection in the real world. First, we can set different thresholds, according to the danger levels of diseases, such as a large threshold for less dangerous diseases and a small threshold for life-threatening diseases. In addition, most of the data in the real world are long time-series data; thus, uncertainties obtained from continuous segments can be inferred jointly. Overall, predicting with rejection is a way to choose more ``certain" predictions.

With the development of deep learning, end-to-end deep learning models have been widely applied to ECG recordings \cite{murat2020application,mathews2018novel,yildirim2018arrhythmia,moskalenko2019deep,li2020deep}. 
Many studies have focused on computer-aided diagnosis using ECG data \cite{zhou2021deep, cai2020ecg, shan2007errors, theodore2003computer, thomson1989computerized}. Automated analysis of ECG data can date back to the 1960s \cite{taback1959digital}, and it was expected that cardiac researchers could be an efficient tool to analyze large-scale data. Maya et al. focused on common errors in computer ECG readings and found that arrhythmia is one of the diseases that most frequent errors are related to \cite{maya2006common}. Bae et al. determined the frequency and nature of erroneous ECG analysis of AF and explained these clinically \cite{bae2012erroneous}. Benefits and limitations still exist, and the automated interpretation of ECGs requires considerable cooperation between clinical experts and CIE manufacturers \cite{jurg2017computer}.

Although our method can reduce misclassifications, it has some limitations. First, because the sampling process of a Bayesian neural network is implemented by dropout, our method only works for deep learning models with dropout components. Second, to calculate the uncertainty, we need to sample multiple times during inference. Therefore, our method has a higher computational complexity than the original model. Third, when dealing with unseen samples, we are not sure how to set the rejection thresholds. High thresholds may cause no use of rejection, while low thresholds may cause undesired silence. Because of these limitations, there is still much work to be done in the future.

In terms of future work, we plan to adopt more general methods to quantify uncertainties, such as deep ensembles \cite{deepensemble}. With sufficient computational resources, a deep ensemble can be a more robust method, and it can be used to quantify uncertainties for models without dropout. At the same time, we can apply distributed and parallel computing, which reduces the computational complexity and facilitates a deep ensemble.

\bibliography{sample}
\bibliographystyle{elsarticle-num-names}

\end{document}